\documentclass{aa}
\usepackage{amssymb}
\usepackage{amsmath}
\usepackage{graphicx}
\usepackage{subfigure}
\usepackage{txfonts}
\usepackage[round]{natbib}
\begin{document}
\title{Very Large Array observations of the 8 o'clock arc lens system: Radio emission and a limit on the star-formation rate}
\author{Filomena Volino\inst{1,}\inst{2}\thanks{Member of the International Max Planck Research School (IMPRS) for Astronomy and Astrophysics at the Universities of Bonn and Cologne.}
\and{Olaf Wucknitz}\inst{1}
\and{John P. McKean}\inst{3}
\and{Michael A. Garrett}\inst{3,}\inst{4,}\inst{5}}
\institute{Argelander-Institut f$\rm \ddot{u}$r Astronomie, Auf dem H$\rm
  \ddot{u}$gel 71, 53121 Bonn, Germany
\and Max Planck Institut f$\rm \ddot{u}$r Radioastronomie, Auf dem H$\rm \ddot{u}$gel 69, 53121 Bonn, Germany
\and ASTRON, Oude Hoogeveensedijk 4, 
7991 PD Dwingeloo, The Netherlands
\and Leiden Observatory, Leiden University, Postbus 9513, 2300 RA Leiden, The Netherlands
\and Centre for Astrophysics and Supercomputing, Swinburne University of
Technology, Australia}
\date{Received  / Accepted}
\abstract
{The 8 o'clock arc is a gravitationally lensed Lyman Break Galaxy (LBG) at redshift $z=2.73$ that has a star-formation rate (SFR) of $\sim270$~M$_{\odot}$~yr$^{-1}$ (derived from optical and near-infrared spectroscopy). Taking the magnification of the system ($\sim 12$) and the SFR into account, the expected flux density of any associated radio emission at 1.4 GHz is predicted to be just 0.1 mJy. However, the lens system is found to be coincident with a radio source detected in the NRAO Very Large Array (VLA) Sky Survey with a flux density of $\sim5$~mJy. If this flux density is attributed to the lensed LBG then it would imply a SFR $\sim$ 11\,000~M$_{\odot}$~yr$^{-1}$, in contrast with the optical and near-infrared derived value.} {We want to investigate the radio properties of this system, and independently determine the SFR for the LBG from its lensed radio emission.}{We have carried out new high resolution imaging with the VLA ain A and B-configurations at 1.4 and 5~GHz.}{We find that the radio emission is dominated by a radio-loud AGN associated with the lensing galaxy. The radio-jet from the AGN partially covers the lensed arc of the LBG, and we do not detect any radio emission from the unobscured region of the arc down to a $3\,\sigma$ flux-density limit of 108~$\mu$Jy~beam$^{-1}$.}{Using the radio data, we place a limit of $\leq750$~M$_{\odot}$~yr$^{-1}$ for the SFR of the LBG, which is consistent with the results from the optical and near-infrared spectroscopy. We expect that the sensitivity of the Expanded VLA will be sufficient to detect many high redshift LBGs that are gravitationally lensed after only a few hours of observing time. The high angular resolution provided by the EVLA will also allow detailed studies of the lensed galaxies and determine if there is radio emission from the lens.}
\keywords{galaxies - high redshift - gravitational lensing}
\titlerunning{Radio observations of the 8 o'clock arc}
\authorrunning{Filomena Volino {et al.}}
\maketitle
\section{Introduction}
Lyman Break Galaxies (LBGs) belong to a population of high redshift objects whose general properties [star-formation rates (SFRs), space density and mass] suggest that they are the progenitors of present day luminous ellipticals and star-forming galaxies. Their spectra, characterized by a blue ultraviolet (UV) continuum, lines from massive stars, weak Ly${\alpha}$ emission, strong interstellar absorption, and dust extinction, are extremely similar to those of nearby star-forming galaxies  \citep{staidel96}.
They have been detected using colour-selection criteria that exploit the Lyman discontinuity in the UV part of the rest-frame spectral energy distribution. For galaxies at higher redshifts, the absorption due to neutral hydrogen is shifted towards longer wavelengths. This results in this population becoming hidden when observations go towards shorter wavelengths \citep{burgarella}. Studies of these objects usually concentrate on their global properties because of their very faint emission. \citet{shapley01} report on a survey of z~$\sim 3$ LBGs, and find that they are forming stars at $\sim 30$~M$_{\odot}$~yr$^{-1}$.  They are also characterized by a wide variety of morphological properties. In general, these galaxies are not classified in terms of Hubble types because of the difficulty in identifying their structural components \citep{giav02}.

More detailed investigations of these early
episodes of star-formation require the additional magnification provided by
gravitational lensing. Although the surface brightness of a lensed galaxy
is conserved, the magnification due to the lensing effect increases its observed integrated flux-density, which can help make detections of distant objects possible. Unfortunately, an advantageous strong lensing geometry is very
rare (lensing probabilities are $\sim$~10$^{-3}$), and there are only a few cases where studies of an LBG boosted by
gravitational lensing have been carried out; for example, MS$1512-$cB$58$ \citep{yee96,pettini00,siana08}, LBG J$213512.73$$-$$010143$
\citep{Smail07,Coppin07}. Also, \citet{LBGII} and \citet{Coppin07} have shown that the dynamics of LBGs can be resolved by gravitational lensing, allowing for detailed kinematic studies of normal galaxies at high redshift.

Unbiased studies of these dust obscured star-forming galaxies are provided by observations in the far-infrared (FIR) to mm bands, where the total stellar and gas content of these dusty environments can be directly probed by measuring the thermal emission of the dust. On the other hand, at radio wavelengths, the non-thermal emission traces recent massive star-formation activity, which gives us another unobscured view of the dust embedded star-forming galaxies. Based on the FIR-radio correlation and a SFR of $\sim 30$~M$_{\odot}$~yr$^{-1}$ \citep{shapley01}, a non-thermal 1.4 GHz luminosity of $\sim 10^{22}$~WHz$^{-1}$ is expected for these galaxies, which implies a sub-mJy flux density. Stacking techniques \citep{ivison07,Carilli08} or deep imaging are therefore necessary to detect these objects when there is no magnification from a gravitational lens. In addition to the technical improvements provided by new facilities at cm wavelengths, gravitational lensing is a powerful tool to investigate LBGs at high redshift. \citet{garrett05} started exploiting this tool using clusters of galaxies as lenses. With this work, we extend the method to individual galaxies as lenses.

The 8 o'clock arc system was first identified by \citet{allam04} while imaging the data of the
Sloan Digital Sky Survey Data Release 4, and follow up spectroscopic
observations confirmed its lensing nature \citep{allam06}. From the time of its discovery the
authors named the system the 8 o'clock arc. The two components of the system are
SDSS J002240.91$+$1431110.4, a luminous red galaxy (LRG), and SDSS
J002240.78$+$143113.9 a very blue and elongated arc (hereafter referred to as the LBG). The LRG is at redshift ${\rm z = 0.38}$ and acts as the lens. The lensed arc of the LBG subtends $9.6 ''$,
and consists of three components: A1, A2 and A3 (with $i= 20.13$, $20.11$ and $20.21$, respectively\footnote{\citealt{allam06} obtained exposures using the SDSS \textit{gri} filters.}; \citealt{allam06}). The SPIcam\footnote{SPIcam CCD imager is mounted on the Astrophysical Research Consortium 3.5 m telescope at the Apache Point Observatory.} \textit{g--}band image of \citet{allam06} also shows a faint fourth component, identified as the counterimage, 5 arcseconds away from the main arc and on the opposite side of the lens galaxy. The redshift of the arc was measured to
be ${\rm z = 2.73}$ \citep{allam06,8oclockIR}. Even taking into account the
lensing magnification of $\approx 12.3$ (from the lens model of \citealt{allam06}), the arc is 2.6 mag more luminous ($\approx $ a factor of 11 in luminosity) than $L_{*}$
for LBGs (where $L_{*}$ is the characteristic Schechter luminosity for LBGs; \citealt{staidel99}). This suggests that the system is going through a
vigorous process of star formation. For this system \citet{allam06} estimated a SFR~$\sim 230$~M$_{\odot}~$yr$^{-1}$ using the relation given in \citet{pettini00} scaled to MS1512$-$cB58. From their optical and near-infrared (NIR) studies, \citet{8oclockIR} found a more robust dust-corrected and de-lensed SFR of $266 \pm 74$~M$_{\odot}~$yr$^{-1}$. Their result confirms that this system is undergoing a very active process of star formation, and shows that the SFR is higher than $\sim 85\%$ of the high redshift LBGs studied by \citet{shapley01,shapley05}.

In the radio, the 8 o'clock arc lens system is coincident with an NVSS\footnote{The National Radio Astronomy Observatory Very Large Array Sky Survey.}  radio source with a 1.4 GHz flux-density of $\sim 5$~mJy (45 arcsecond resolution; \citealt{nvss}). Such a large flux-density at 1.4~GHz would imply a huge SFR $\sim$ 11\,000~M$_\odot$~yr$^{-1}$, which would contradict the estimates from the optical and NIR spectroscopy. Taking the optical and NIR derived SFR of $\sim 270$~M$_{\odot}~$yr$^{-1}$ and a total magnification of $\mu \sim 12$, the gravitationally lensed 1.4 GHz flux density for the LBG is expected to be just  $\sim 0.12~$mJy.

In this paper, we present the results from new Very Large Array (VLA) observations of the 8 o'clock arc lens system at 1.4 and 5 GHz, which provide a better and more clear understanding of its radio properties. From these observations, we confirm the 1.4 GHz flux-density of the system is $\sim$ 5~mJy, but our higher resolution observations show that most of the emission is due to an AGN within the foreground lens. From our data we measure an upper limit for the continuum radio emission of the LBG and calculate an upper limit for the SFR.

For all calculations, we made use of a cosmology with H$_{0} = 70 $ km\,s$^{-1}~$Mpc$^{-1}$, $\rm{\Omega_{m} = 0.3}$ and $\rm{\Omega_{\lambda} = 0.7}$.

\section{Very Large Array observations of the 8 o'clock arc system}
\label{obss}
\subsection{Observations}
\label{obs}
We observed the 8 o'clock arc in November 2007 using the VLA in B-configuration at 1.4 GHz [synthesized half power beam width (HPBW) of ~$\sim 3.9~$arcsec] and at 5 GHz (synthesized HPBW $\sim$ 1.2~arcsec), and again at 1.4 GHz one year later, when the telescope was in
A-configuration (HPBW $ \sim$ 1.4~arcsec). The higher resolution and more sensitive 1.4 GHz data (second observing run) were necessary to disentangle the weak radio emission of the lensed LBG from the lens.  For the first observing run, the total time on source was 6 hours, divided equally between the two observing frequencies, while one year later we observed the system for 7 hours. In order to reduce the potential effect of bandwidth smearing, at 1.4 GHz we used the correlator in
spectral line mode, using 25 MHz of bandwidth and 3.125 MHz wide channels. At this frequency a larger field of view is required for imaging (in order to remove the response of unwanted sources), and therefore the use of a narrower bandwidth is required to reduce chromatic aberrations which cannot be neglected. In addition to this, continuum observations in spectral line mode are usual  preferred at this observing frequency as narrow interferences, which can be a problem at this or longer wavelength, can be identified. For both runs, 3C\,48 and $0010+174$ were used as the flux-scale and gain (amplitude and phase) calibrators, respectively. The data were taken in 2 IFs, but for both 1.4 GHz observations, one of the IFs was corrupted, and therefore was not used in imaging. For the first run (November 2007), at 5 GHz we used a switching cycle of 1.5 and 5 min between the calibrator (0010$+$174) and the lens system, while at 1.4 GHz we used 1.5 and 20 min; in the second run (November 2008), the array configuration was larger (A-array), therefore we used a cycle-time of 1.5 and 10 min in order to compensate for the less stable phases. 

The data reduction was performed using the AIPS (Astronomical Imaging Processing Software) package, provided by the
National Radio Astronomy Observatory. The calibration strategy was the following: the flux-scale was set using 3C\,48; afterwards amplitude and  phase calibration was performed using 0010$+$174. In both cases, for the 1.4 GHz data, bandpass calibration was necessary in order to identify variations of amplitudes and phases across the band; 3C\,48 was used for this. This step was not required for the 5 GHz data which were taken in continuum mode.
At 1.4 GHz the primary beam has a half-width at half maximum of $\sim 18$ arcminutes while at 5 GHz it is $\sim$ 5 arcminutes, therefore wide field imaging techniques and deep cleaning were necessary in order to map all of the sources in the field-of-view, and remove their interfering sidelobes from the region of interest around the lens system. At both frequencies the whole primary beam was mapped.

In order to reach a high accuracy in the mapping of the extended emission from the system, a \textit{multi scale clean} approach was used. This technique is powerful when mapping weak extended emission over a large area which contributes to the total flux but has a low signal-to-noise ratio at full resolution \nolinebreak \citep{multiscale1,multiscale2,multiscale3}.

\begin{figure*}
\centering
\subfigure[]{\label{av307}\includegraphics[width=9cm]{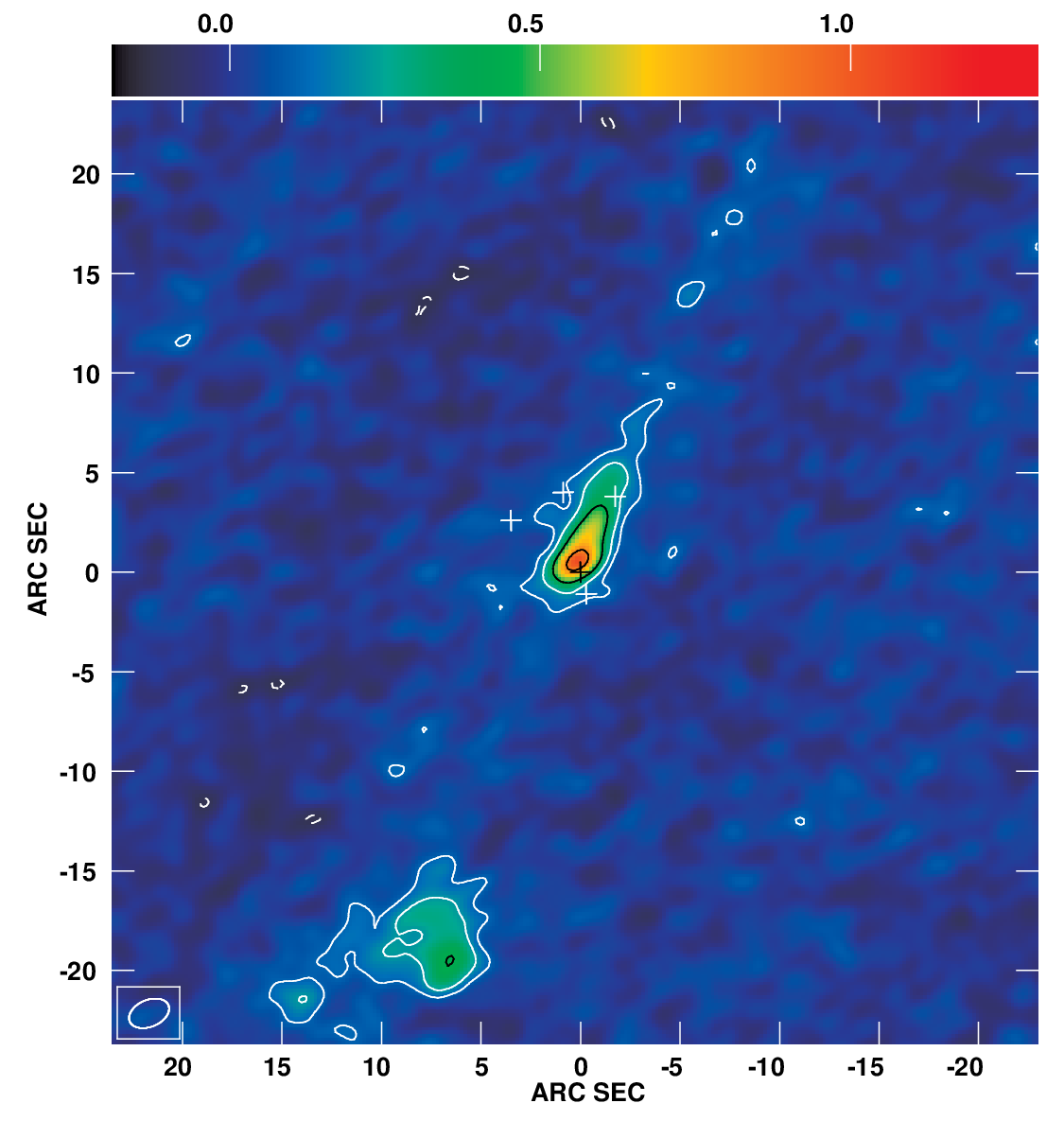}}
\centering
\subfigure[]{\label{aw717C}\includegraphics[width = 9cm]{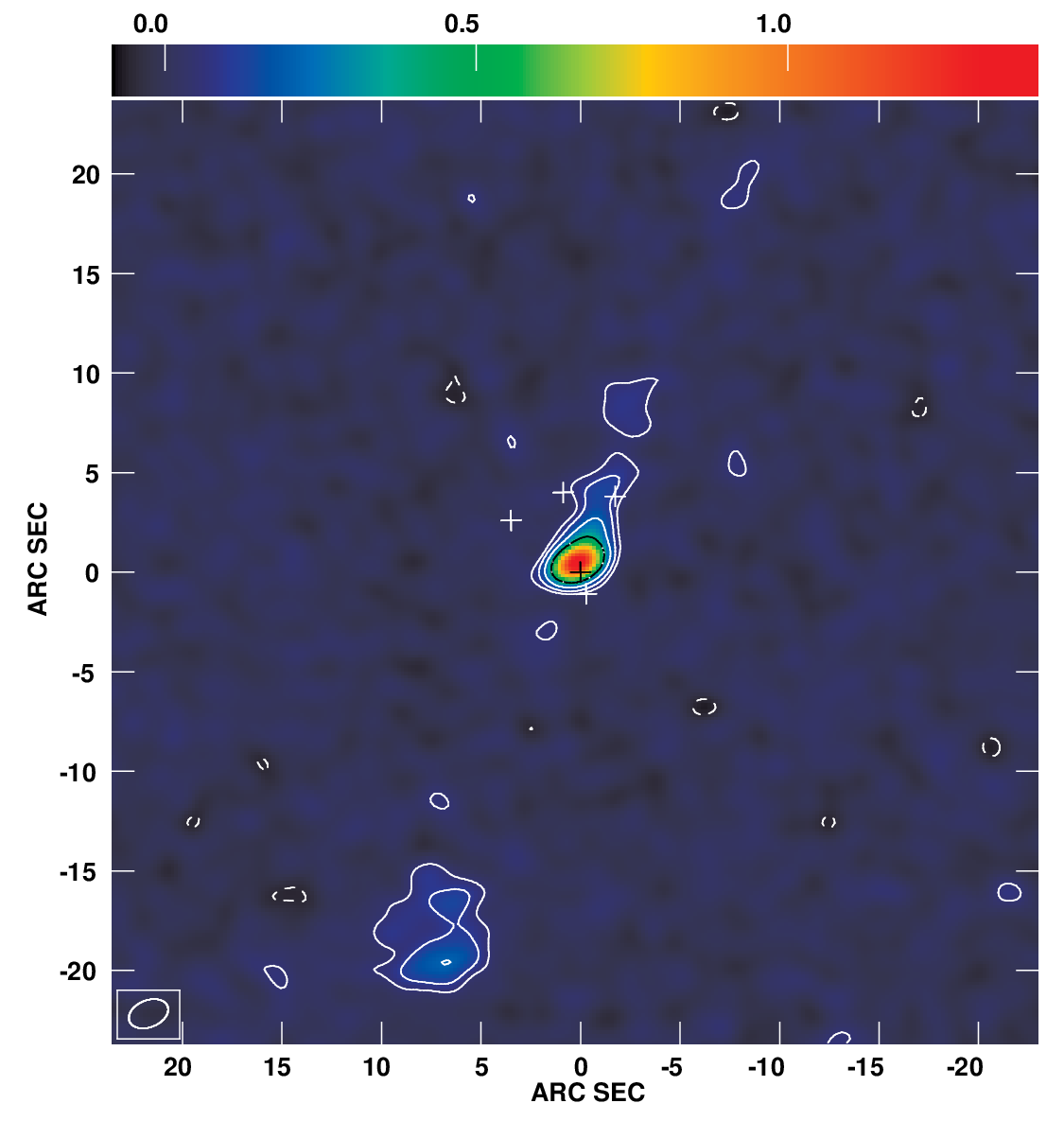}}
\caption{The 8 o'clock arc gravitational lens system at 1.4 and 5 GHz, as observed with the VLA. We find no firm detection of radio emission from the lensed arc, but do find that there is a radio-loud AGN associated with the lensing galaxy. In both images the crosses identify the lensed arc components and the lens galaxy. (a) The 1.4 GHz image has been restored with $2.1\times1.4$ arcsec  beam (position angle of $-$67$^{\circ}$). The contours are shown at $(-3, 3, 6, 12, 24) \times 36~\mu$Jy, the rms map noise. (b) The 5 GHz map has a restoring beam of $2.1\times1.4$ arcsec (position angle of $-$68.1$^{\circ}$). The contours are shown at $(-3, 3, 6, 12, 24) \times 17~\mu$Jy, the rms map noise. The theoretical rms of the 1.4 and 5 GHz maps are 25 and $15~\mu$Jy~beam$^{-1}$, respectively. The colour-scale is in mJy~beam$^{-1}$. North is up and east is left.}
\label{vla}
\end{figure*}

\begin{figure}
\centering
\includegraphics[angle=-90,width=9cm,clip]{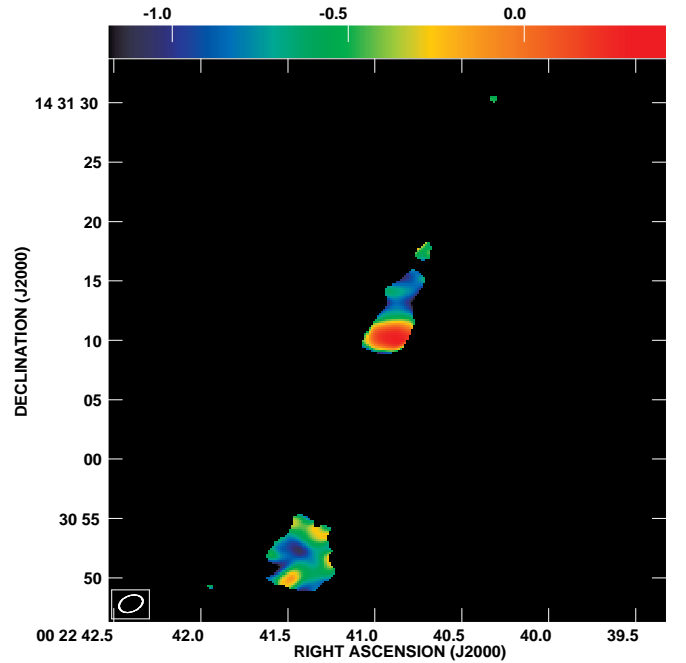}
\centering
\caption{The spectral index map of the 8 o'clock arc lens system between 1.4 and 5 GHz, where S($\nu$)~$\propto \nu^{\alpha}$. This map has been made using only the emission detected at $\geq 3\,\sigma$ in the 1.4 and 5 GHz images. The colour-scale is the spectral index.}
\label{spec}
\end{figure}

\begin{figure}
\centering
\includegraphics[angle=0,width=9cm,clip]{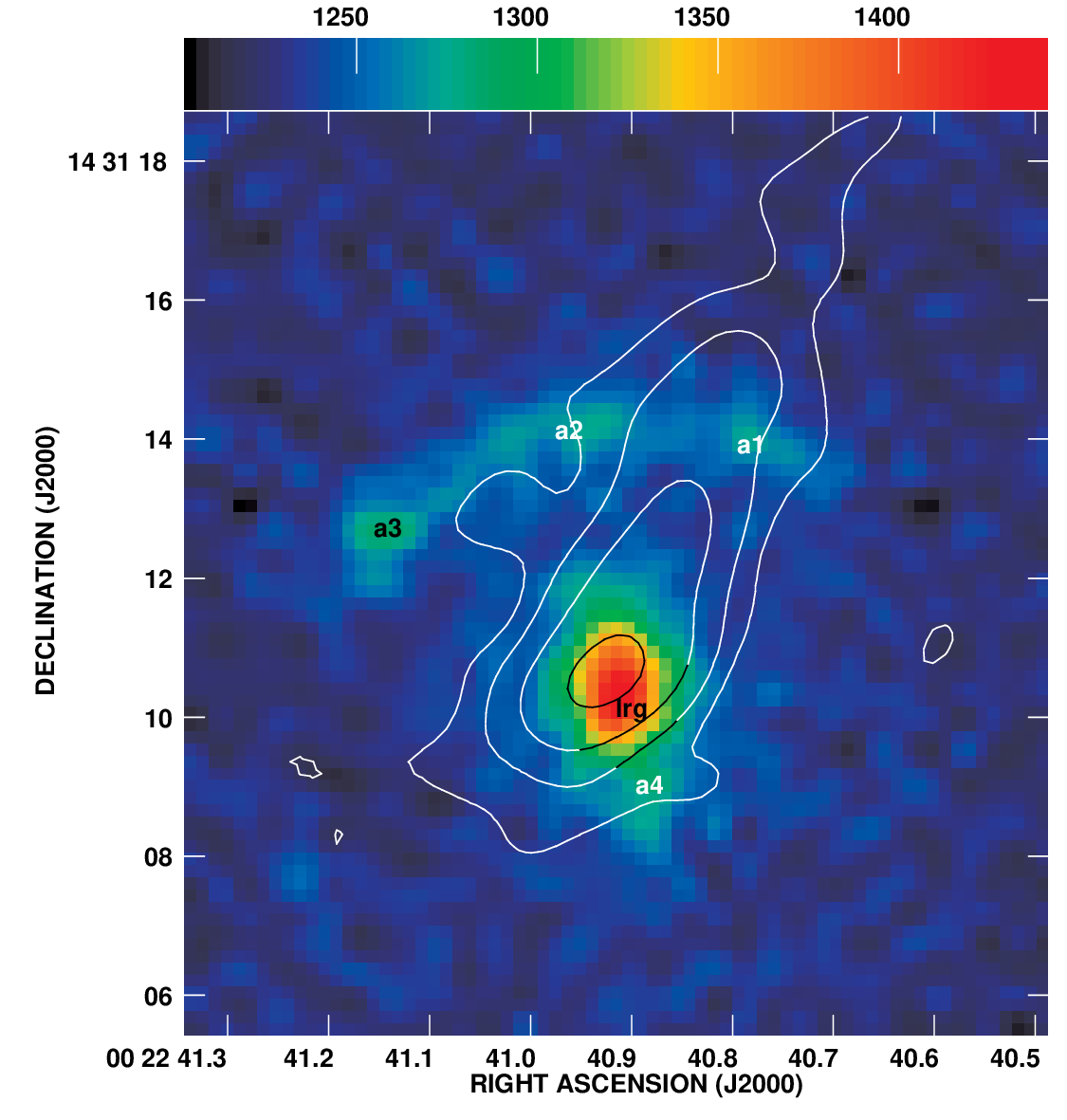}
\centering
\caption{The optical image of the 8 o'clock arc lens system with the radio contours of the 1.4 GHz observations overlayed. The contours are shown at $(-3, 3, 6, 12, 24) \times 36~\mu$Jy as in fig. \ref{av307}. No radio emission from the arc is detected above the $3\,\sigma$ confidence contours. The four lensed images are labelled a1, a2, a3 (arc) and a4 (counter image). The lensing galaxy is labelled lrg.}
\label{optical}
\end{figure}

\subsection{Data analysis}
\label{analysis}
In Figure \ref{vla}, we show the 1.4 and 5 GHz images of the 8 o'clock arc system that were taken with the VLA. The crosses indicate the positions of the lensed images and the lens galaxy. As already mentioned, in the NVSS catalogue the lens system is identified with a radio source that has a 1.4 GHz flux-density of $\sim 5$~mJy at 45 arcsec resolution \citep{nvss}. The observations presented here better resolve the radio emission of the system and we find that the lens is dominated by the emission from an AGN. The peak of the radio emission is coincident with the optical position of the lens galaxy. In Figure \ref{spec} we show the spectral index map of the system between 1.4 and 5 GHz. We find that the peak in the radio emission has a spectral index of $\alpha \sim +0.25$, where $S_{\nu} \propto \nu^\alpha$, which is consistent with a synchrotron self-absorbed radio source. Therefore, we associate the brightest radio component, which is coincident with the optical position of the lensing galaxy, as the compact core of the AGN. At both frequencies, we see that there is radio emission that propagates from the core region towards the north-west. This radio emission has a steeper spectral index of $\alpha \sim -0.80$, consistent with an optically-thin radio jet. This would suggest that the AGN has a morphology that is consistent with an FR\,I type radio galaxy. Extended emission is also seen opposite to the jet direction with respect to the AGN core (towards the south-east), which could be from a radio lobe; the steep spectral index ($\alpha \sim -0.70$) and morphology agree with this classification. The integrated flux density of the system at 1.4 GHz is $\sim 5$~mJy, in agreement with the NVSS flux density.

In Figure \ref{optical}, we show an overlay of the 1.4 GHz radio contours and the SDSS {\it i}-band image. We find that the radio jet covers about two thirds of the arc, contaminating the radio emission from components A1 and A2 of the LBG. However, component A3 of the arc is relatively free from the emission of the radio jet and can be used to estimate the SFR of the LBG. The lensed 1.4 GHz flux-density at the position of A3 is $80~\pm~36~\mu$Jy~beam$^{-1}$, which is a $2.2\,\sigma$ detection, and therefore, is not statistically significant. Based on the rms noise in our VLA map, we place a firm $3\,\sigma$ upper limit of $108~\mu$Jy~beam$^{-1}$  for the 1.4 GHz flux-density of component A3. We have also measured the flux-density of component A2 and the total flux-density of the arc (A1 + A2 + A3), but these data are contaminated by the emission from the radio-jet. All our measurements are presented in Table~\ref{table}.

\subsection{A limit on the SFR of the LBG from the radio data}
\label{arc}

Galaxies with large SFRs tend to show increased supernovae activity. The winds and shocks from these supernovae accelerate free electrons, which produce synchrotron emission that is observable at radio wavelengths. This results in a tight correlation between the radio emission and the rate massive stars ($\geq 5$~M$_{\odot}$) are expected to form in a galaxy \citep{condon92}. To make a fair comparison between the results from the optical/NIR and the radio data for the 8 o'clock arc we must measure the SFR using the same initial mass function \citep{salpeter55} and down to the same stellar mass limit ($\geq0.1$~M$_{\odot}$). This needs the relation between the SFR and the radio emission to be modified. This was done by \citealt{yun01} and found to be,
\begin{equation}
\label{conv}
\frac{\rm SFR}{\rm (M_{\odot}~yr^{-1})} = \left( 5.9 \pm 1.8 \right)  \times 10^{-22}~\frac{L_{\rm 1.4~GHz}}{\rm(W~Hz^{-1})}.
\end{equation}
The rest-frame luminosity at 1.4 GHz is calculated using,
\begin{equation}
\label{lumm}
L_{\rm 1.4~GHz} = 4 \pi D_{L}^2 S_{\rm 1.4~GHz} (1+z)^{-(1 + \alpha)}
\end{equation}
where $D_L$ is the luminosity distance, $\alpha$ is the spectral index of the LBG (assumed to be $-0.8$) and $S_{\rm 1.4~GHz}$ is the observed monochromatic flux density. The factor ${\rm (1+z)^{-(1+ \alpha)}}$ accounts for the $k$-correction \citep{schmidt86}.


To calculate the SFR of the LBG we must also correct the observed flux-density for the lensing magnification. The lensing model constructed by \citet{allam06} gives a total magnification of 12.3 for the arc, with each component (A1, A2 and A3) having a magnification of about 4. It is likely that there could be differential magnification across the full extent of the arc given its large size. However, since we have at best an upper-limit for the radio emission from the arc and that the resolution of our radio data matches the size of the individual components, we have adopted the magnifications found by \citet{allam06} in our calculations. This gives also consistency when comparing the radio-derived SFR and the optical one. The estimated unlensed luminosity  and SFR of the LBG is given in Table~\ref{table}.

\begin{table*}
\begin{center}
\begin{minipage}[t]{\columnwidth}
\caption{A summary of the radio emission observed from the 8 o'clock arc lens system. From our VLA observations we quote the observed (gravitationally lensed) flux-densities and the rest-frame (unlensed) luminosities at 1.4 GHz. We give the estimated SFRs, corrected for the lensing magnification. For comparison, we also show the data from NVSS and the expected radio emission based on the optical/NIR derived SFR.}
\label{table}
\centering
\renewcommand{\footnoterule}{}  
\begin{tabular}{lcccc}
\hline \hline
 Component						&$\mu$~$^{a}$ & $S_{\rm 1.4~GHz}$ & $L_{\rm 1.4~GHz}$ &  SFR   \\ 
  						& 	&($\mu$Jy) 		&(W~Hz$^{-1}$)		&(${\rm M_{\odot}~yr^{-1}}$)\\
\hline
A3 ($2.2\,\sigma$ detection)	& 4	& $80\pm36$~$^{b}$		& $9\pm4 \times 10^{23}$		& $560\pm300$	\\
A3 ($3\,\sigma$ upper limit)	& 4	& $\leq108$~$^{b}$		& $\leq1.3 \times 10^{24}$	& $\leq750$		\\
A2 (including jet)			& 4	& $95\pm36$~$^{b}$		& $1.1\pm0.4 \times 10^{24}$	& $660\pm310$	\\
A1+A2+A3 (including jet)		& 12	& $430\pm70$		& $1.7\pm0.3 \times 10^{24}$	& $1000\pm350$	\\
All (NVSS)				& 12 & $4700\pm500$	& $1.9\pm0.2 \times 10^{25}$	& $11000\pm3500$	\\
\hline \hline
\\
\multicolumn{5}{c}{Prediction from optical/NIR-based SFR } \\
\\
\hline \hline
\multicolumn{2}{c}{SFR} &\multicolumn{1}{c}{$S_{\rm 1.4~GHz}$} & \multicolumn{2}{c}{$L_{\rm 1.4~GHz}$}\\
\multicolumn{2}{c}{(${\rm M_{\odot}~yr^{-1}}$)} &\multicolumn{1}{c}{($\mu$Jy)} & \multicolumn{2}{c}{(W~Hz$^{-1}$)}\\
\hline
\multicolumn{2}{c}{$270\pm$75} &\multicolumn{1}{c}{$120\pm50$}& \multicolumn{2}{c}{$4.6\pm1.9 \times 10^{23}$}  \\   \hline \hline
\\ 
~$^{a}$ magnification correction \\
~$^{b}$ flux density per beam
\end{tabular}
\end{minipage}
\end{center}
\end{table*}

\section{Discussion}
\label{xxx}

\subsection{Radio emission from the 8 o'clock arc}

We have observed the 8 o'clock arc gravitational lens system to investigate the source of the excess radio emission at 1.4 GHz. Previous observations from NVSS found a radio source with a flux-density of $\sim5$~mJy. If this flux density is associated with the lensed LBG, the implied SFR of $\sim11000$~M$_{\odot}$~yr$^{-1}$ is in conflict with the optical/NIR derived value of $\sim270$~M$_{\odot}$~yr$^{-1}$. Three possible explanations could account for this; i) there is a radio-loud AGN within the lensing galaxy, ii) there is a radio-loud AGN associated with the lensed LBG and iii) there is a genuine discrepancy between the optical/NIR and radio methods for determining the SFR. Our observations with the VLA have shown that there is a radio-loud AGN within the lensing galaxy that dominates the radio emission from the 8 o'clock arc gravitational lens system. In principle, our observations should have been sensitive enough to detect the radio emission from the LBG as the predicted flux-density from the arc is 120~$\mu$Jy. Due to losing half of the data from a corrupted IF and the radio jet from the AGN covering most of the arc, we could only place a limit on the SFR from the uncontaminated region of the lensed arc. Our limit of $\leq750$~M$_{\odot}$~yr$^{-1}$  ($3\,\sigma$) is consistent with the SFR found from the optical/NIR data.

We were unfortunate that this system has a radio-loud lensing galaxy whose emission covered part of the LBG; from surveys of lens systems at radio wavelengths about 1 in 10 lensing galaxies, which are typically massive early-types, are radio-loud. For example, the Cosmic Lens All-Sky Survey (CLASS; \citealt{browne03,myers03}) found 22 gravitational lenses and two systems had radio-loud lensing galaxies (B2108+213; \citealt{mckean05,more08}, and B2045+265; \citealt{fassnacht99,mckean07}).

\subsection{Radio emission from other LBGs and future prospects}

\citet{Carilli08} showed the first robust statistical detection of sub-$\mu$Jy radio emission for a sample of high redshift LBGs from the COSMOS field at z~$\sim 3$. Their results found an average flux density of ${\rm 0.90~\pm~0.21~\mu Jy }$ at 1.4 GHz, which implies a total SFR of $\simeq 31$~M$_{\odot}$~yr$^{-1}$, based on the radio--FIR correlation for low redshift star-forming regions. The comparison of this result with those obtained from UV data, for star-forming galaxies at the same redshift, showed a discrepancy between the SFR implied by the non-thermal radio luminosity and that derived from the UV spectra. In particular, the authors find that the ratio of radio- to UV-based SFRs is 1.8, indicating either a smaller dust attenuation factor (standard values for LBGs are $\sim$ 5) or an attenuation of the radio luminosity to SFR conversion factor at z $\sim3$. Discriminating between these two possibilities requires a deeper look into the interaction between CMB photons and relativistic electrons at such redshifts (increased electron cooling due to Inverse Compton scattering off the CMB; \citealt{Carilli08}), and into the properties of LBGs. The radio-FIR correlation may as well require a deeper insight as the radio-derived SFRs rely on the assumption that this correlation is as tight as for local galaxies. A discrepancy between UV and radio-derived SFR is also found by \citet{ivison07}, in their 1.4 GHz VLA survey of starburst galaxies up to z~$\leq 1.3$.

For lensed systems, Spitzer IR spectroscopy and photometry of two  LBGs, namely, MS $1512-$cB58 \citep{siana08} and the LBG J$213512.73-010143$  (also known as the Cosmic Eye; \citealt{siana09}) show that the UV spectral slope overpredicts the reddening by dust and thus the SFR measurements. In particular, for the system MS$1512-$cB$58$ ($z=2.73$), \citet{siana08} find a SFR of $\sim 20 - 40$ M$_{\odot}$~yr$^{-1}$, consistent with the SFR derived from the dust corrected  H${\alpha}$ luminosity, but 5 times lower than the UV-derived SFR. For the  Cosmic Eye (${\rm z=3.074}$),  \citet{siana09} find that the SFR inferred from the IR luminosity is 8 times lower than that predicted from the rest-frame UV properties. In addition, for the Cosmic Eye, CO studies indicate a SFR $\sim 60$ M$_{\odot}$~yr$^{-1}$  \citep{Coppin07}. These numbers imply that calculating a radio continuum dust unbiased SFR for a sample of star-forming galaxies could help resolve the conflicting SFRs that have been found from IR and UV studies.

The new generation of radio facilities, such as the Expanded Very Large Array (EVLA), will have significantly better surface brightness sensitivity due to improved receivers and larger instantaneous bandwidths. L-band observations with 1 GHz bandwidth using the EVLA will reach sensitivities of $\sim8~\mu$Jy in 1 hour, which corresponds to a SFR of $\sim220$~M$_{\odot}$~yr$^{-1}$ at redshift 2.7. To reach the average SFR for LBGs of $\sim30$~M$_{\odot}$~yr$^{-1}$ \citep{shapley01,shapley05} will require of order 60 hours of integration. Clearly, studies of lensed LBGs will also benefit from the increased sensitivity. The added advantage is that the lensing magnification will allow more systems to be observed over a shorter amount of time and give higher-resolution imaging of the structure of these galaxies. There has recently been a large increase in the numbers of lensed LBGs being found making the prospects of detailed studies of meaningful samples of these galaxies possible in the near future.


\section{Conclusion}
\label{summ}
With this work we reported the study at radio frequencies of the  8 o'clock arc gravitational lens system, which from optical/NIR data is expected to have a large SFR of $\sim270$~M$_{\odot}$~yr$^{-1}$. Through observations at 1.4 and 5 GHz with the VLA, the lens galaxy was found to host an AGN that has a radio morphology consistent with an FR\,I-type radio galaxy. We confirm a total flux-density of  $\sim5$~mJy for this system at 1.4~GHz, which was first measured from NVSS. We conclude that the main contribution to the flux density of the system is from the lensing galaxy. We measure an upper limit for the radio emission of the lensed LBG to be  $108~\mu$Jy~beam$^{-1}$ ($3\,\sigma$). This value is consistent with what is predicted by the optical/NIR derived SFR for this system. The improved sensitivity provided by the Expanded VLA will allow better and more significative detections of the emission from this and other lensed LBGs in the future, opening a new insight into this high-redshift star-forming galaxy population.

\section*{Acknowledgments}
The National Radio Astronomy Observatory is a facility of the National Science Foundation operated under cooperative agreement by Associated Universities, Inc. This work was supported by the Emmy-Noether Programme of the Deutsche Forschungsgemeinschaft, reference WU588/1-1. FV was supported for part of this research through a stipend from the International Max Planck Research School (IMPRS) for Astronomy and Astrophysics at the Universities of Bonn and Cologne. FV thanks Richard Porcas and Eduardo Ros for helpful and inciting conversations regarding the interpretation of the faint radio emission from the 8 o'clock arc. We thank the anonymous referee who made invaluable comments that greatly improved the manuscript.
{
\bibsep0.4ex
\bibliographystyle{aa}
\bibliography{biblio}

\begin{thebibliography}{33}
\expandafter\ifx\csname natexlab\endcsname\relax\def\natexlab#1{#1}\fi

\bibitem[{{Allam} {et~al.}(2007){Allam}, {Tucker}, {Lin}, {Diehl}, {Annis},
  {Buckley-Geer}, \& {Frieman}}]{allam06}
{Allam}, S.~S., {Tucker}, D.~L., {Lin}, H., {et~al.} 2007, \apjl, 662, L51

\bibitem[{{Allam} {et~al.}(2004){Allam}, {Tucker}, {Smith}, {Lee}, {Annis},
  {Lin}, {Karachentsev}, \& {Laubscher}}]{allam04}
{Allam}, S.~S., {Tucker}, D.~L., {Smith}, J.~A., {et~al.} 2004, \aj, 127, 1883

\bibitem[{{Bhatnagar} \& {Cornwell}(2004)}]{multiscale1}
{Bhatnagar}, S. \& {Cornwell}, T.~J. 2004, \aap, 426, 747

\bibitem[{{Browne} {et~al.}(2003){Browne}, {Wilkinson}, {Jackson}, {Myers},
  {Fassnacht}, {Koopmans}, {Marlow}, {Norbury}, {Rusin}, {Sykes}, {Biggs},
  {Blandford}, {de Bruyn}, {Chae}, {Helbig}, {King}, {McKean}, {Pearson},
  {Phillips}, {Readhead}, {Xanthopoulos}, \& {York}}]{browne03}
{Browne}, I.~W.~A., {Wilkinson}, P.~N., {Jackson}, N.~J.~F., {et~al.} 2003,
  \mnras, 341, 13

\bibitem[{{Burgarella} {et~al.}(2006){Burgarella}, {P{\'e}rez-Gonz{\'a}lez},
  {Tyler}, {Rieke}, {Buat}, {Takeuchi}, {Lauger}, {Arnouts}, {Ilbert},
  {Barlow}, {Bianchi}, {Lee}, {Madore}, {Malina}, {Szalay}, \&
  {Yi}}]{burgarella}
{Burgarella}, D., {P{\'e}rez-Gonz{\'a}lez}, P.~G., {Tyler}, K.~D., {et~al.}
  2006, \aap, 450, 69

\bibitem[{{Carilli} {et~al.}(2008){Carilli}, {Lee}, {Capak}, {Schinnerer},
  {Lee}, {McCraken}, {Yun}, {Scoville}, {Smol{\v c}i{\'c}}, {Giavalisco},
  {Datta}, {Taniguchi}, \& {Urry}}]{Carilli08}
{Carilli}, C.~L., {Lee}, N., {Capak}, P., {et~al.} 2008, \apj, 689, 883

\bibitem[{{Condon}(1992)}]{condon92}
{Condon}, J.~J. 1992, \araa, 30, 575

\bibitem[{{Condon} {et~al.}(1998){Condon}, {Cotton}, {Greisen}, {Yin},
  {Perley}, {Taylor}, \& {Broderick}}]{nvss}
{Condon}, J.~J., {Cotton}, W.~D., {Greisen}, E.~W., {et~al.} 1998, \aj, 115,
  1693

\bibitem[{{Coppin} {et~al.}(2007){Coppin}, {Swinbank}, {Neri}, {Cox}, {Smail},
  {Ellis}, {Geach}, {Siana}, {Teplitz}, {Dye}, {Kneib}, {Edge}, \&
  {Richard}}]{Coppin07}
{Coppin}, K.~E.~K., {Swinbank}, A.~M., {Neri}, R., {et~al.} 2007, \apj, 665,
  936

\bibitem[{{Cornwell}(2008)}]{multiscale2}
{Cornwell}, T.~J. 2008, IEEE Journal of Selected Topics in Signal Processing,
  vol.~2, issue 5, pp.~793-801, 2, 793

\bibitem[{{Fassnacht} {et~al.}(1999){Fassnacht}, {Blandford}, {Cohen},
  {Matthews}, {Pearson}, {Readhead}, {Womble}, {Myers}, {Browne}, {Jackson},
  {Marlow}, {Wilkinson}, {Koopmans}, {de Bruyn}, {Schilizzi}, {Bremer}, \&
  {Miley}}]{fassnacht99}
{Fassnacht}, C.~D., {Blandford}, R.~D., {Cohen}, J.~G., {et~al.} 1999, \aj,
  117, 658

\bibitem[{{Finkelstein} {et~al.}(2009){Finkelstein}, {Papovich}, {Rudnick},
  {Egami}, {Le Floc'h}, {Rieke}, {Rigby}, \& {Willmer}}]{8oclockIR}
{Finkelstein}, S.~L., {Papovich}, C., {Rudnick}, G., {et~al.} 2009, \apj, 700,
  376

\bibitem[{{Garrett} {et~al.}(2005){Garrett}, {Knudsen}, \& {van der
  Werf}}]{garrett05}
{Garrett}, M.~A., {Knudsen}, K.~K., \& {van der Werf}, P.~P. 2005, \aap, 431,
  L21

\bibitem[{{Giavalisco}(2002)}]{giav02}
{Giavalisco}, M. 2002, \araa, 40, 579

\bibitem[{{Greisen} {et~al.}(2009){Greisen}, {Spekkens}, \& {van
  Moorsel}}]{multiscale3}
{Greisen}, E.~W., {Spekkens}, K., \& {van Moorsel}, G.~A. 2009, \aj, 137, 4718

\bibitem[{{Ivison} {et~al.}(2007){Ivison}, {Chapman}, {Faber}, {Smail},
  {Biggs}, {Conselice}, {Wilson}, {Salim}, {Huang}, \& {Willner}}]{ivison07}
{Ivison}, R.~J., {Chapman}, S.~C., {Faber}, S.~M., {et~al.} 2007, \apjl, 660,
  L77

\bibitem[{{McKean} {et~al.}(2005){McKean}, {Browne}, {Jackson}, {Koopmans},
  {Norbury}, {Treu}, {York}, {Biggs}, {Blandford}, {de Bruyn}, {Fassnacht},
  {Mao}, {Myers}, {Pearson}, {Phillips}, {Readhead}, {Rusin}, \&
  {Wilkinson}}]{mckean05}
{McKean}, J.~P., {Browne}, I.~W.~A., {Jackson}, N.~J., {et~al.} 2005, \mnras,
  356, 1009

\bibitem[{{McKean} {et~al.}(2007){McKean}, {Koopmans}, {Flack}, {Fassnacht},
  {Thompson}, {Matthews}, {Blandford}, {Readhead}, \& {Soifer}}]{mckean07}
{McKean}, J.~P., {Koopmans}, L.~V.~E., {Flack}, C.~E., {et~al.} 2007, \mnras,
  378, 109

\bibitem[{{More} {et~al.}(2008){More}, {McKean}, {Muxlow}, {Porcas},
  {Fassnacht}, \& {Koopmans}}]{more08}
{More}, A., {McKean}, J.~P., {Muxlow}, T.~W.~B., {et~al.} 2008, \mnras, 384,
  1701

\bibitem[{{Myers} {et~al.}(2003){Myers}, {Jackson}, {Browne}, {de Bruyn},
  {Pearson}, {Readhead}, {Wilkinson}, {Biggs}, {Blandford}, {Fassnacht},
  {Koopmans}, {Marlow}, {McKean}, {Norbury}, {Phillips}, {Rusin}, {Shepherd},
  \& {Sykes}}]{myers03}
{Myers}, S.~T., {Jackson}, N.~J., {Browne}, I.~W.~A., {et~al.} 2003, \mnras,
  341, 1

\bibitem[{{Nesvadba} {et~al.}(2006){Nesvadba}, {Lehnert}, {Eisenhauer},
  {Genzel}, {Seitz}, {Davies}, {Saglia}, {Lutz}, {Tacconi}, {Bender}, \&
  {Abuter}}]{LBGII}
{Nesvadba}, N.~P.~H., {Lehnert}, M.~D., {Eisenhauer}, F., {et~al.} 2006, \apj,
  650, 661

\bibitem[{{Pettini} {et~al.}(2000){Pettini}, {Steidel}, {Adelberger},
  {Dickinson}, \& {Giavalisco}}]{pettini00}
{Pettini}, M., {Steidel}, C.~C., {Adelberger}, K.~L., {Dickinson}, M., \&
  {Giavalisco}, M. 2000, \apj, 528, 96

\bibitem[{{Salpeter}(1955)}]{salpeter55}
{Salpeter}, E.~E. 1955, \apj, 121, 161

\bibitem[{{Schmidt} \& {Green}(1986)}]{schmidt86}
{Schmidt}, M. \& {Green}, R.~F. 1986, \apj, 305, 68

\bibitem[{{Shapley} {et~al.}(2001){Shapley}, {Steidel}, {Adelberger},
  {Dickinson}, {Giavalisco}, \& {Pettini}}]{shapley01}
{Shapley}, A.~E., {Steidel}, C.~C., {Adelberger}, K.~L., {et~al.} 2001, \apj,
  562, 95

\bibitem[{{Shapley} {et~al.}(2005){Shapley}, {Steidel}, {Erb}, {Reddy},
  {Adelberger}, {Pettini}, {Barmby}, \& {Huang}}]{shapley05}
{Shapley}, A.~E., {Steidel}, C.~C., {Erb}, D.~K., {et~al.} 2005, \apj, 626, 698

\bibitem[{{Siana} {et~al.}(2009){Siana}, {Smail}, {Swinbank}, {Richard},
  {Teplitz}, {Coppin}, {Ellis}, {Stark}, {Kneib}, \& {Edge}}]{siana09}
{Siana}, B., {Smail}, I., {Swinbank}, A.~M., {et~al.} 2009, \apj, 698, 1273

\bibitem[{{Siana} {et~al.}(2008){Siana}, {Teplitz}, {Chary}, {Colbert}, \&
  {Frayer}}]{siana08}
{Siana}, B., {Teplitz}, H.~I., {Chary}, R.-R., {Colbert}, J., \& {Frayer},
  D.~T. 2008, \apj, 689, 59

\bibitem[{{Smail} {et~al.}(2007){Smail}, {Swinbank}, {Richard}, {Ebeling},
  {Kneib}, {Edge}, {Stark}, {Ellis}, {Dye}, {Smith}, \& {Mullis}}]{Smail07}
{Smail}, I., {Swinbank}, A.~M., {Richard}, J., {et~al.} 2007, \apjl, 654, L33

\bibitem[{{Steidel} {et~al.}(1999){Steidel}, {Adelberger}, {Giavalisco},
  {Dickinson}, \& {Pettini}}]{staidel99}
{Steidel}, C.~C., {Adelberger}, K.~L., {Giavalisco}, M., {Dickinson}, M., \&
  {Pettini}, M. 1999, \apj, 519, 1

\bibitem[{{Steidel} {et~al.}(1996){Steidel}, {Giavalisco}, {Pettini},
  {Dickinson}, \& {Adelberger}}]{staidel96}
{Steidel}, C.~C., {Giavalisco}, M., {Pettini}, M., {Dickinson}, M., \&
  {Adelberger}, K.~L. 1996, \apjl, 462, L17+

\bibitem[{{Yee} {et~al.}(1996){Yee}, {Ellingson}, {Bechtold}, {Carlberg}, \&
  {Cuillandre}}]{yee96}
{Yee}, H.~K.~C., {Ellingson}, E., {Bechtold}, J., {Carlberg}, R.~G., \&
  {Cuillandre}, J. 1996, \aj, 111, 1783

\bibitem[{{Yun} {et~al.}(2001){Yun}, {Reddy}, \& {Condon}}]{yun01}
{Yun}, M.~S., {Reddy}, N.~A., \& {Condon}, J.~J. 2001, \apj, 554, 803

\end{thebibliography}
}

\end{document}